\documentclass[11pt]{article} 
\usepackage{graphicx}
\newcommand{\be}{\begin{equation}}
\newcommand{\ee}{\end{equation}}
\newcommand{\bea}{\begin{eqnarray}}
\newcommand{\eea}{\end{eqnarray}}
\newcommand{\sn}{{\rm sn}}

\newcommand{\dn}{{\rm dn}}
\newcommand{\cn}{{\rm cn}}

\begin{document}

\vspace{0.5in}
\begin{center}
{\Large{\bf Solitons of the Symmetric $\phi^4$-$\phi^2 |\phi|$-$\phi^2$ Triple Well Model}}
\end{center}

\begin{center}
{\Large{\bf Avinash Khare}}\\
{Physics Department, Savitribai Phule Pune University, Pune 411007, India}
\end{center}
	
\begin{center}
{\Large{\bf Avadh Saxena}}\\
{Theoretical Division and Center for Nonlinear Studies,\\
Los Alamos National Laboratory, Los Alamos, New Mexico 87545, USA}
\end{center}
\vspace{0.9in}
{\bf Abstract}

A symmetric $\phi^4$-$\phi^2 |\phi|$-$\phi^2$ model has recently attracted a lot of 
attention due to its usefulness in studying tunable phase transitions. We analyze the 
behavior of this model for the entire range of parameters and obtain its kink and pulse 
solutions. For completeness, we also present several periodic solutions of this model. 
Furthermore, we present a generalized symmetric $\phi^{4n}$-$\phi^{2n}|\phi|$-$\phi^2$ 
model where $n = 1, 2, 3, ...$ and obtain its kink and pulse solutions for arbitrary $n$.


\section{Introduction}
 
Recently, a symmetric $\phi^4$-$\phi^2 |\phi|$-$\phi^2 $ triple well model containing a third  
order term has attracted considerable attention in the literature \cite{halperin, fumika}.  
Specifically, a weakly first-order phase transition character in certain superfluids and 
superconductors can be captured by such a model \cite{halperin}.  This model also 
serves as a testbed to study topological defect formation in phase transitions with 
tunable order \cite{fumika} within the context of the Kibble-Zurek mechanism 
\cite{kibble,zurek}. 

However, to our surprise, there does not seem to exist a detailed analysis of the 
various possible phases as well as  topological and non-topological defects admitted 
by this model. This is our main motivation and one of the tasks that we have undertaken  
in this paper. In particular, we discuss in detail the behavior of the model
\be\label{1}
V(\phi) = \frac{a}{2} \phi^2 - \frac{|c|}{3} \phi^2 |\phi| 
+ \frac{b}{4} \phi^4 + d\,,
\ee
where $d$ is a constant which one can choose as required, and in general $a$, $b$, 
$c$ can be positive or negative.  For $c=0$ and $b>0$ it reduces to the standard 
double well $\phi^4$ model ($\phi^4$-$\phi^2$) for second order transitions. The value 
of $c$ as compared to $a$ for $b>0$ determines the strength of the first order character 
of the transition. Thus, this model is suitable to study tunable order phase transitions \cite{fumika}. 
Note that, this is another model for first order transitions in addition to the standard 
triple well $\phi^6$ model ($\phi^6$-$\phi^4$-$\phi^2$) \cite{sanati1} and the asymmetric 
double well $\phi^4$ model \cite{sanati2, sanati3}.  The latter arises when a third order nonlinearity  
is symmetry allowed (e.g., $\phi^4$-$\phi^3$-$\phi^2$) \cite{sanati2, sanati3}.  Note that there 
is no modulus in the third order term in this case unlike the model considered in this paper.  

The plan of the paper is as follows. In Sec. II we discuss the behavior of
this model for various possible values of the parameters $a, b, c$. In
Sec. III we obtain several novel hyperbolic pulse and kink solutions of the 
model in the most interesting case of $a, b, c > 0$. In Sec. IV we generalize 
the symmetric $\phi^4$-$\phi^2 |\phi|$-$\phi^2$ model by presenting a 
one-parameter family of symmetric triple well $\phi^{4n}$-$\phi^{2n}|\phi|$-$\phi^2$ 
models, where $n = 1,2,3,...$, and 
\be\label{2a}
V(\phi) = \frac{a}{2} \phi^2 - \frac{|c|}{(2n+1)} \phi^{2n} |\phi| 
+ \frac{b}{4n} \phi^{4n} + d\,.
\ee
We then discuss the behavior of the generalized model for various possible values 
of the parameters $a, b, c$. Further, for an arbitrary integer $n$, we obtain its hyperbolic 
kink and pulse solutions in the most interesting case of $a,b,c > 0$. In Sec.~V we 
summarize the new results obtained in this paper and point out some of the open 
problems. For completeness, in the Appendix we present a few representative periodic 
solutions of the above symmetric $\phi^4$-$\phi^2 |\phi|$-$\phi^2$ model.

\section{Analysis of the Symmetric Triple Well Model}

Consider the symmetric triple well model characterized by the potential
(\ref{1}). Let us discuss the behavior of this model for different possible
values of the parameters $a, b, c$. 

{\bf Case I: $a, b > 0$}

Let us first consider the most interesting case of $a, b > 0$. On
differentiating Eq. (\ref{1}) we obtain
\be\label{2}
\frac{dV}{d\phi} =  \phi (a - |c| |\phi| + b |\phi|^2)\,.
\ee
Thus the extrema of the potential are at 
\be\label{3}
\phi = 0\,,~~ |\phi| = \frac{|c| \pm \sqrt{c^2-4ab}}{2b}\,.
\ee
Therefore, if $c^2 < 4ab$ then the potential has minima only at $\phi = 0$. On the
other hand, if $4ab < c^2 < (9/2) ab$, $\phi = 0$ is still an absolute minimum 
while $|\phi| =|\phi_{-}| = \frac{|c| + \sqrt{c^2-4ab}}{2b}$ is a local minimum. 
The two minima at $\phi = 0$ and $|\phi| = |\phi_{-}|$ are degenerate with 
each other provided $2c^2 = 9ab$. Note that since it is $|\phi_{-}|$,  
one actually has three degenerate minima at $\phi = 0$ and $\phi = \pm |\phi_{-}|$. 
Finally, for $2c^2 > 9ab$, $\phi = 0$ is a local minimum and $|\phi| = |\phi_{-}|$ 
is an absolute minimum. Note also that for $2c^2 > 9 ab$, one has a maximum
at $\phi = |\phi_{+}| = \frac{|c| - \sqrt{c^2-4ab}}{2b}$. As we will see below,
the local minimum at $\phi = 0$ disappears at $a = 0$ and so also the two
maxima, and $\phi = 0$ is now a point of inflection while $\phi = |\phi_{-}|$
continues to be an absolute minimum. Finally, for $a < 0$, now $\phi = 0$ is 
a maximum while $|\phi_{-}|$ continues to be a minimum. Notice that this 
picture is very similar to that of the $a \phi^2$-$c \phi^4$-$b\phi^6$ model 
\cite{sanati1}.

{\bf Case II: $b, a < 0$}

In this case the extrema of the potential are at 
\be\label{4}
\phi = 0\,,~~ |\phi| = |\phi_{-}| = \frac{|c| + \sqrt{c^2+4|a|b}}{2b}\,.
\ee
Out of these while $\phi = 0$ is a maximum, $|\phi| = \phi_{-}$ is a
minimum.

{\bf Case III: $b > 0, a = 0$}

In this case the extrema of the potential are at 
\be\label{5}
\phi = 0\,,~~ |\phi| = |\phi_{-}| = \frac{|c|}{b}\,.
\ee
Out of these while $\phi = 0$ is a point of inflection, $|\phi| = |c|/b$
is an absolute minimum. Thus, there must be a kink solution from 
$\phi = -c/b$ to $\phi = +c/b$. 

{\bf Case IV: $a > 0, b = 0$}

In this case while $\phi = 0$ is an absolute minimum, $|\phi| = a/|c|$ 
corresponds to maxima. Hence there should be a pulse solution around 
$\phi = 0$ going from $\phi = -a/c$ to $\phi = + a/c$. 

{\bf Case V: $a, b, c < 0$}

Note, instead of $|c|$, we will consider $-|c|$. Thus, the extrema of the 
potential are at 
\be\label{6}
\phi = 0\,,~~ |\phi_{\pm}| = \frac{|c| \pm \sqrt{c^2-4|a||b|}}{2|b|}\,.
\ee

It is easy to see that while the potential goes to $-\infty$ as $\phi \rightarrow 
\pm \infty$, for finite $\phi$ one has maxima at $\phi = 0$ and 
$\phi = \phi_{+}$ while a minimum is at $\phi = \phi_{-}$. Thus, one 
could have a pulse solution around $\phi = 0$.

\section{Kink and Pulse Solutions in Case $a, b > 0$}

For the remaining part of the paper we mostly confine ourselves to the most 
interesting case of $a, b > 0$.
For simplicity, we now do a dimensional analysis so that the physics will
essentially depend on one parameter $\mu = {ab}/{c^2}$. If we write the 
Ginzburg-Landau type free energy, it is given by 
\be\label{6d}
F_{T} = (G/2) \left(\frac{d\phi}{dx}\right)^2 + (a/2) \phi^2 - (|c|/3) \phi^2 |\phi|  
+(b/4) \phi^4\,, 
\ee 
where $G$ is the gradient coefficient that sets the scale for the kink or pulse 
width.  On rescaling, that is considering 
\be\label{6e}
F_{R} = \frac{c^4}{b^3}\,,~~\eta = \frac{b\phi}{|c|}\,,
\ee
one finds that
\be\label{6f}
\frac{F_{T}}{F_{R}} = g \left(\frac{d\eta}{dx}\right)^2 + (\mu/2) \eta^2 
- (1/3)\eta^2 |\eta| + (1/4) \eta^4\,,
\ee
where $\mu = {ab}{/c^2}$. From now onwards we consider various 
solutions of the field equation (where for simplicity we revert to denoting $\phi$ 
instead of $\eta$)
\be\label{7}
\phi_{xx} =  \phi^3 - \phi |\phi| + \mu \phi\,.
\ee

In this case, at the point of three degenerate minima (that is, first order 
transition point), i.e. $\mu = 2/9$, one expects two (half) kink and two (half)
antikink solutions. On the other hand, below the transition point, i.e. when 
$\mu < 2/9$, there are two degenerate minima and one expects a kink (and 
the corresponding antikink) solution. Besides, we also expect pulse solutions 
in the cases of both above and below the first order transition point, i.e. for 
$2/9 < \mu < 1/4$ as well as when $\mu < 2/9$. Note, when $2/9 < \mu < 1/4$, 
there are two local minima and $\phi = 0$ is an absolute minimum. In Fig.~1, 
a plot of the potential $V(\phi)$ vs $\phi$, Eq. (\ref{7}), is given for various 
representative values of $\mu$ (note $\mu = {ab}/{c^2}$). 

It is worth pointing out that the symmetric model which we are discussing here 
is very different from the asymmetric $\phi^4$ model \cite{sanati2, sanati3} for 
which instead of the field Eq. (\ref{7}), we have the equation
\be\label{7a}
\phi_{xx} =  \phi^3 - \phi^2  + \mu \phi\,.
\ee
The corresponding potential has an absolute minimum at $\phi = 0$ and a local minimum 
at $\phi = \phi_{+} = \frac{1+\sqrt{1-4\mu}}{2}$ in case $2/9 < \mu < 1/4$ while $\phi = 0$ 
is a local minimum and $\phi = \phi_{+}$ is an absolute minimum in case
$\mu < 2/9$; and $\phi = 0$ and $\phi = \phi_{+}$ are degenerate at 
$\mu = 2/9$. As a result, the physics described by the symmetric and the 
asymmetric $\phi^4$ models is very different. Note that asymmetric double well 
potentials are of significant physical interest in the study of quantum tunneling 
and localization \cite{weiner, rastelli,song}.  

\begin{figure}[h] 
\includegraphics[width=3.0 in]{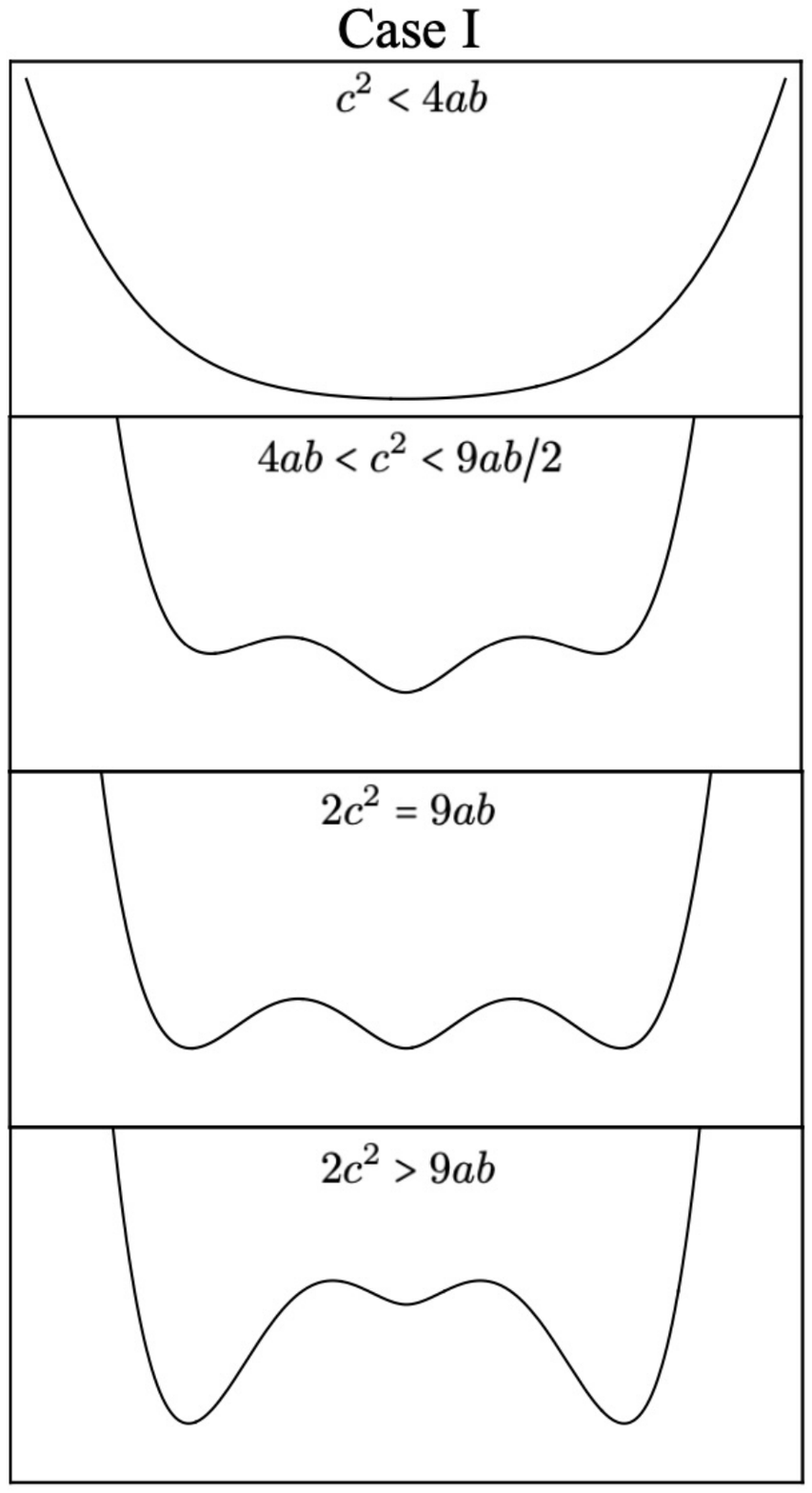}
\caption{The shape of the potential $V(\phi)$ vs $\phi$, Eq. (\ref{7}), centered around 
$\phi=0$ for various parameter values. (a) Top panel: Well above the 
transition temperature (or pressure) $T_c$.  (b) Just above $T_c$. (c) Three degenerate 
minima at $T_c$. (d) Bottom panel: Two degenerate minima below $T_c$.}
\end{figure} 

\subsection{Kink and Antikink Solutions at $\mu = 2/9$}

We now show that at the point $\mu = 2/9$ where there are three 
degenerate minima at $\phi = 0$ and $\phi = \pm \phi_{-} 
= \frac{|c|+\sqrt{c^2-4ab}}{2b}$, there are several exact solutions
including the kink and the periodic kink solutions. While we discuss the 
kink solution below, the corresponding periodic kink solution is discussed in 
the Appendix. 

Instead of the rescaled field Eq.~(\ref{7}), we consider the following 
field equation
\be\label{8}
\phi_{xx} =  \phi^3 -  \phi^2 + \mu \phi
\ee
and obtain those solutions for which $\phi = |\phi|$, so that they are
automatically also the solutions of the field Eq. (\ref{7}). 

{\bf Kink Solutions at $\mu = 2/9$}

It is easy to check that
\be\label{9}
\phi = A + B \tanh(\beta x)\,,
\ee
is an exact solution of Eq. (\ref{8}) provided
\be\label{10}
A = \frac{1}{3}\,,~~\beta^2 = \frac{1}{18}\,,~~B = \pm \frac{1}{3}\,.
\ee
Clearly for this solution $\phi = |\phi|$ and hence Eq. (\ref{9}) is also an 
exact solution of Eq. (\ref{7}). Observe that since $B = \pm A$, hence
we have  both the kink and the antikink solutions for $\phi > 0$ as well as for
$\phi < 0$. For $\phi > 0$ we have
\be\label{11}
\phi(x) = \frac{1}{3} [1 \pm \tanh(\beta x)]\,,
\ee
while for $\phi < 0$ we have
\be\label{12}
\phi(x) = -\frac{1}{3}[1 \pm \tanh(\beta x)]\,,
\ee
where $\beta = \sqrt{{1}/3\sqrt{2}}$. In Fig.~2 a plot of the kink and the 
corresponding antikink solutions $\phi(x)$ vs $x$, Eqs. (\ref{11}) and 
(\ref{12}), is given at the transition temperature $T_c$ corresponding to 
$\mu = 2/9$.

\begin{figure}[h] 
\includegraphics[width=5.0 in]{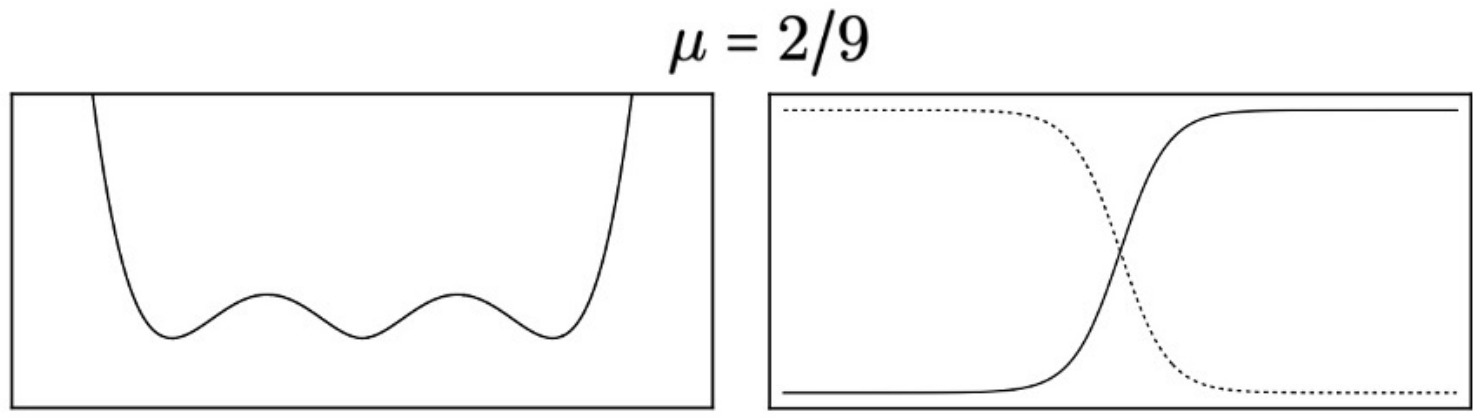}
\caption{Right panel: Kink and antikink profiles $\phi(x)$ vs $x$, Eqs. (\ref{11}) 
and Eq. (\ref{12}), at the transition temperature $T_c$ corresponding to 
the potential $V(\phi)$ (left panel) with three degenerate minima at $\mu = 2/9$. } 
\end{figure} 

\subsection{Kink and Antikink Solutions for $\mu < 2/9$}

For $\mu < 2/9$, we have two degenerate absolute minima at $\phi = \pm 
\phi_{-}$ and hence there should be a kink and an antikink solution. In this 
case the potential can be reexpressed as
\be\label{13}
V(\phi) = \frac{1}{4} (\delta -|\phi|)^2 \left[|\phi|^2 +2\left(\delta-\frac{2}{3}\right)
|\phi| +\delta\left(\delta -\frac{2}{3}\right)\right]\,,
\ee
where
\be\label{14}
\delta = \phi_{-} = \frac{1+\sqrt{1-4\mu}}{2}\,.
\ee

In order to get the kink solution, we need to solve the first order
self-dual equation
\be\label{15}
\frac{d\phi}{dx} = \sqrt{2V(\phi)}\,,
\ee
which in our case takes the form
\be\label{16}
\int \frac{d\phi}{\sqrt{|\phi|^2+2(\delta -\frac{2}{3})|\phi|
+\delta(\delta -\frac{2c}{3b})}~(\delta -|\phi|)} 
= \frac{x}{\sqrt{2}}\,.
\ee
We can breakup the left hand side into two parts depending on whether $\phi > 0$
or $\phi < 0$. Thus we have to solve the following first order equations
depending on if $\phi > 0$ or $\phi < 0$:
\be\label{17}
\int \frac{d\phi}{\sqrt{\phi^2+2(\delta -\frac{2}{3})\phi
+\delta(\delta -\frac{2}{3})}~(\delta -\phi)} 
= \frac{x}{\sqrt{2}}\,,~~~\phi > 0\,,
\ee
\be\label{18}
\int \frac{d\phi}{\sqrt{\phi^2-2(\delta -\frac{2}{3})\phi
+\delta(\delta -\frac{2}{3})}~(\delta +\phi)} 
= \frac{x}{\sqrt{2}}\,,~~~\phi < 0\,. 
\ee
The integrals in Eqs. (\ref{17}) and (\ref{18}) are easily done using
\be\label{19}
\int \frac{dx}{x\sqrt{C x^2 + Bx +A}} = -\frac{1}{\sqrt{A}}
\tanh^{-1}\left[\frac{2A+Bx}{2\sqrt{A}\sqrt{A+Bx+Cx^2}}\right]\,,
\ee
provided $A > 0$. For $\phi > 0$, on making the substitution $u = \delta -\phi$,
the integral (\ref{17}) takes the form
\be\label{20}
-\int \frac{du}{u \sqrt{u^2-4(\delta -\frac{1}{3})u
+2\delta(2\delta -1)}} = \frac{x}{\sqrt{2}}\,.
\ee
On using the integral (\ref{20}) we then obtain for $\phi > 0$
\bea\label{21}
\tanh\left[\sqrt{\delta(2\delta-1)} x\right] 
= \frac{2\delta(2\delta-1)-2(\delta -\frac{1}{3})u}
{\sqrt{2\delta(2\delta-1)}\sqrt{u^2 -4(\delta-\frac{1}{3})u
+2\delta(2\delta-1)}}\,.
\eea
Note that we have used $u = \delta - \phi$. From here it immediately follows 
that
\be\label{22}
\lim_{x \rightarrow \infty}  \phi = \delta\,.
\ee
On the other hand if $\phi = 0$, then  $x > 0$ since in that case
\be\label{23}
\tanh\left[\sqrt{\delta(2\delta-1)} x_0\right] = 
\frac{[\sqrt{1-4\mu}-\frac{1}{3}]^{1/2}}{\sqrt{1-4\mu}} > 0\,,
\ee
since $\mu < 2/9$. 

Proceeding in the same way, for $\phi < 0$ we find that
\bea\label{21a}
&&\tanh\left[\sqrt{\delta(2\delta-1)} x\right] 
\nonumber \\
&&=-\frac{2\delta(2\delta-1)-2(\delta -\frac{1}{3})u}
{\sqrt{2\delta(2\delta-1)}\sqrt{u^2 -4(\delta-\frac{1}{3})u
+2\delta(2\delta-1)}}\,.
\eea
Note that here $u = \delta + \phi$. It then follows that
\be\label{22a}
\lim_{x \rightarrow -\infty} \phi = -\delta\,.
\ee
On the other hand if $\phi = 0$, then  $x < 0$ since in that case
\be\label{23a}
\tanh\left[\sqrt{\delta(2\delta-1)} x_0\right] = 
-\frac{[\sqrt{1-4\mu}-\frac{1}{3}]^{1/2}}{\sqrt{1-4\mu}} < 0\,,
\ee
Note that in general 
$\phi(-\infty) = -\phi(\infty)$. Thus, while for $\phi > 0$, $\phi$ goes from 
$\phi = \delta$ to $\phi = 0$ in case $x$ goes from $\infty$ to $x_0$ as given 
by Eq. (\ref{21}), while for $\phi < 0$, $\phi$ goes from $\phi = -\delta$ to 
$\phi = 0$ in case $x$ goes from $-\infty$ to $-x_0$ as given by 
Eq. (\ref{21a}). In Fig.~3, a plot of the kink solution $\phi(x)$ as a function
of $x$ is given corresponding to $T < T_c$ for a representative value of 
$\mu = 0.0979$. 

From the above analysis as well as from the figure it is clear that the region
$-x_0$ to $x_0$ is not accessible to the kink solution. One possible 
explanation is that the kink profile is like the trajectory of a ball that hits
the wall at $x_0$ and bounces because of the $|\phi|$ term. A similar ball on the 
other side of $-x_0$ does the same bounce. Thus the two pieces from $-\infty$ 
to $-x_0$ and $x_0$ to $\infty$ can be considered as the kink solution. 
Clearly one needs to have a better understanding of this kink solution.

\begin{figure}[h] 
\includegraphics[width=5.0 in]{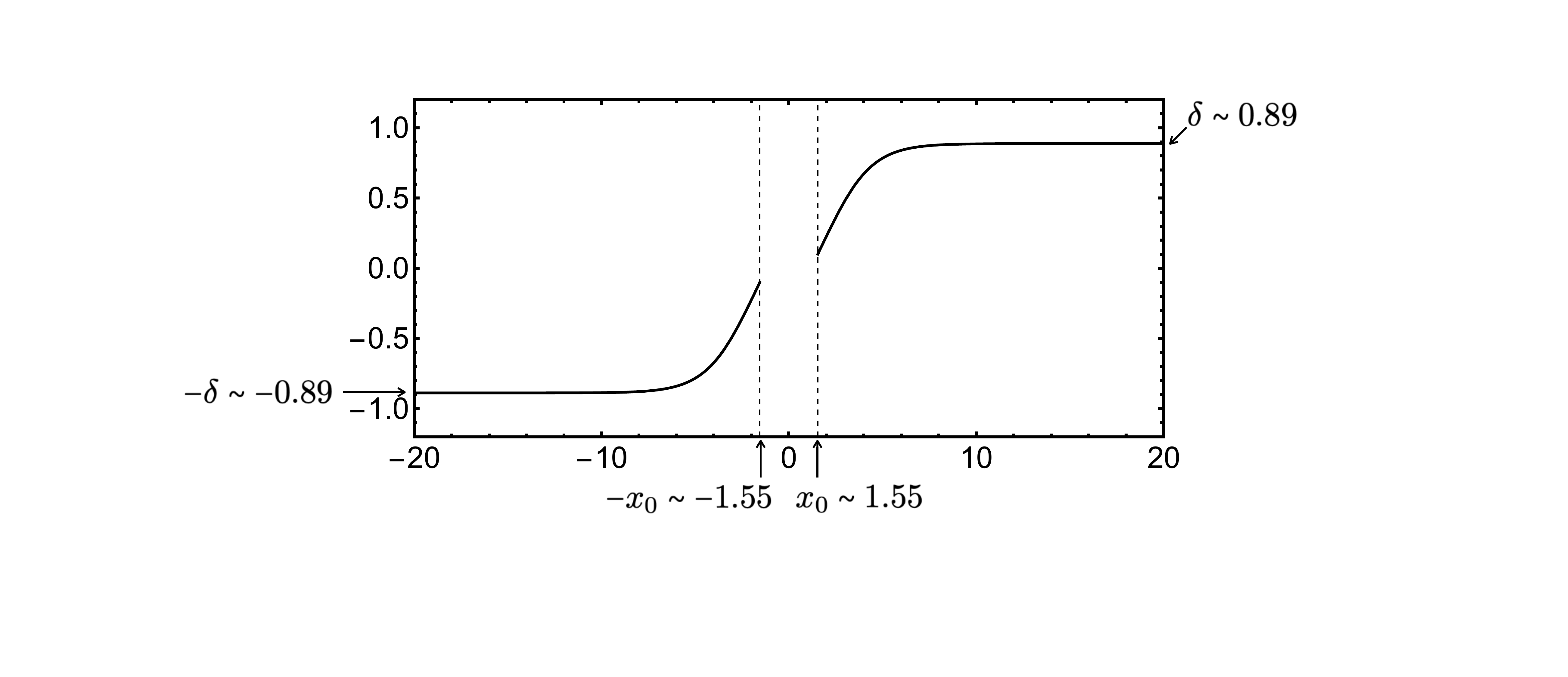}
\caption{Kink solution $\phi(x)$ vs $x$, Eqs. (\ref{21}) and (\ref{21a}), 
corresponding to $T < T_c$ for a representative value of $\mu = 0.0979$. 
The region $-x_0<x<x_0$ is inaccessible to the field $\phi(x)$. The 
corresponding potential $V(\phi)$ is given in the left panel of Fig.~4 }
\end{figure}

\subsection{Pulse Solution For $\mu < 2/9$}

The symmetric $\phi^4$ triple well Eq. (\ref{8}) admits a hyperbolic pulse solution
\be\label{1.1}
\phi = \frac{A}{B+\cosh^2(\beta x)}\,,~~B > 0\,,
\ee
provided 
\be\label{1.2}
\mu = 4\beta^2\,,~~\frac{9\mu}{2} = \frac{4B(B+1)}{(2B+1)^2} < 1\,,~~A = 
\frac{3(2B+1)\mu}{2}\,.
\ee
Notice that this solution is only valid if $\mu < 2/9$, i.e. when the potential
has two degenerate minima. Further, for this 
solution $\phi = |\phi|$ and hence Eq. (\ref{1.1}) is also an exact solution of 
Eq. (\ref{7}). 

On using the well known identity \cite{as}
\be\label{1.2a}
\frac{1}{B+\cosh^2(y)} = \frac{\tanh(y+\Delta)-\tanh(y-\Delta)}
{\sinh(2\Delta)}\,,~~B = \sinh^2(\Delta)\,,
\ee
we can reexpress the solution of Eq. (\ref{1.1}) as a 
superposition of a kink-antikink-like solution, i.e.
\be\label{1.3}
\phi =\sqrt{\frac{\mu}{2}} [\tanh(\beta x +\Delta)
-\tanh(\beta x -\Delta)]\,,
\ee
where $B = \sinh^2(\Delta)$. 
A plot of the pulse solution $\phi(x)$ vs $x$, Eq. (\ref{1.1}) , is given in Fig.~4 for
$T < T_c$ at a representative value of $\mu $.
\begin{figure}[h] 

\includegraphics[width=5.5 in]{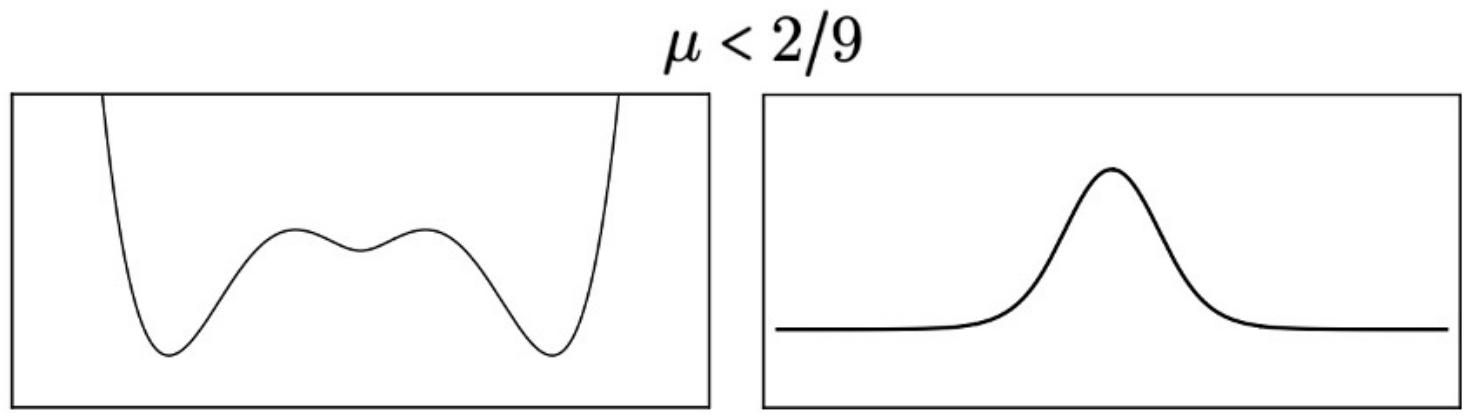}
\caption{Right panel: Pulse solution $\phi(x)$  vs $x$, Eq. (\ref{1.1}), 
centered at $\phi=0$ corresponding to the potential $V(\phi)$ (left panel) at  
$T < T_c$ for a representative value of $\mu < 2/9$. }
\end{figure} 

\subsection{Pulse Solution For $2/9 < \mu < 1/4$}

The symmetric $\phi^4$ triple well Eq. (\ref{8}) also admits another
hyperbolic pulse solution
\be\label{1.4}
\phi = D - \frac{A}{B+\cosh^2(\beta x)}\,,~~B > 0\,,
\ee
provided 
\be\label{1.5} 
D = \mu +  D^2\,,
\ee
\be\label{1.6}
A^2 = 8B(B+1)\beta^2\,,
\ee
\be\label{1.7}
4\beta^2 =  D^2 -\mu\,,
\ee
\be\label{1.8}
(3D -1)A = 6(2B+1) \beta^2\,.
\ee

On solving Eqs. (\ref{1.5}) and (\ref{1.7}) we obtain
\be\label{1.8a}
D = \frac{1+\sqrt{1-4\mu}}{2}\,,~~\beta^2 = \frac{\sqrt{1-4\mu}
[1+\sqrt{1-4\mu}]}{8}\,.
\ee
Further, on solving Eqs. (\ref{1.5}) to (\ref{1.8}) one obtains a quadratic 
equation for the ratio $A/D = y$ in terms of $B$. In particular, we obtain
\be\label{1.9}
2y^2 -3(2B+1)y+4B(B+1) = 0\,,
\ee
which on solving gives
\be\label{1.10}
y = \frac{A}{D} = \frac{3(2B+1) \pm \sqrt{(2B+1)^2 +8}}{4}\,.
\ee
Thus, solution (\ref{1.4}) can be reexpressed as
\be\label{1.11}
\phi = \frac{D[(B-y)+\cosh^2(\beta x)]}{B+\cosh^2(\beta x)}\,.
\ee
We must demand that $\phi = |\phi|$ which is ensured if $B+1-y > 0$
since $\cosh^2(\beta y) \ge 1$. Clearly, this is ensured if we 
choose the lower root of $y$ in Eq. (\ref{1.10}). Thus, we must choose
\be\label{1.12}
y = \frac{A}{D} = \frac{3(2B+1) - \sqrt{(2B+1)^2 +8}}{4}\,.
\ee
One can now show that
\be\label{1.13}
\frac{1}{\mu} = \frac{2y[(2B+1)-y]^2}{B(B+1)[3(2B+1)-4y]}\,,
\ee
where $y$ is given in terms of $B$ by Eq. (\ref{1.12}). It is then
easily shown that for this solution $2/9 < \mu < 1/4$.
Since for this solution $\phi = |\phi|$, hence Eq. (\ref{1.4}) is also 
an exact solution of Eq. (\ref{7}).  Thus, this solution is valid when 
there are two local minima and an absolute minimum at $\phi = 0$.

On using the identity of Eq. (\ref{1.2a})
one can reexpress the solution (\ref{1.4}) as a superposition of 
kink-antikink-like solution over a background, i.e.
\be\label{1.14}
\phi = D - \sqrt{2}\beta [\tanh(\beta x +\Delta) -\tanh(\beta x -\Delta)]\,,
\ee
where $B = \sinh^2(\Delta)$ while $D$ and $\beta$ are given by Eq. (\ref{1.8a}).
A plot of the pulse solution $\phi(x)$ vs $x$, Eq. (\ref{1.4}), is given in 
Fig.~5 for $T > T_c$ at a representative value of $2/9 < \mu < 1/4$.

\begin{figure}[h] 
\includegraphics[width=5.0 in]{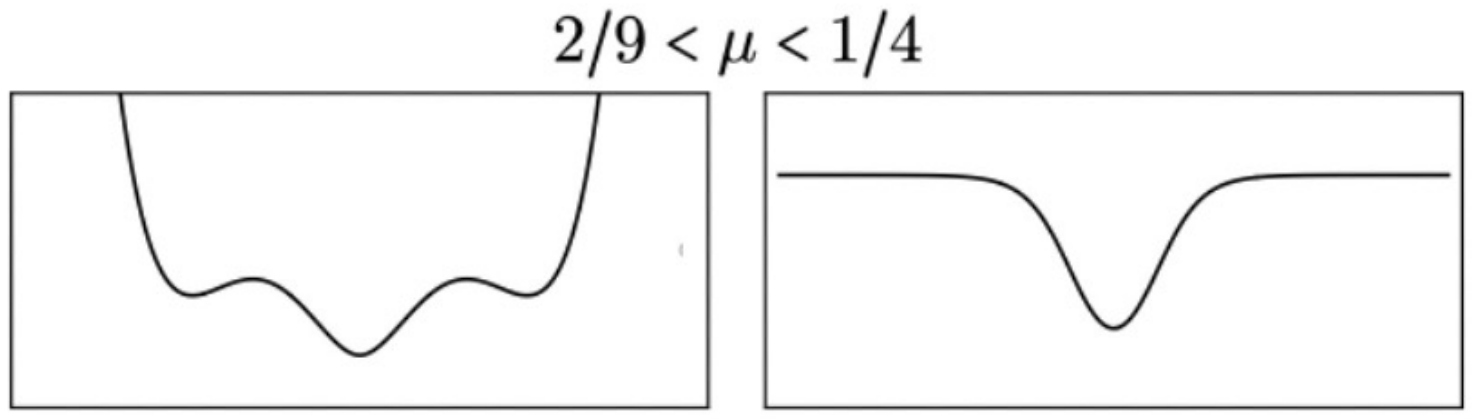}
\caption{Right panel: Pulse solution $\phi(x)$ vs $x$, Eq. (\ref{1.4}), centered at 
$\phi=0$ corresponding to the potential $V(\phi)$ (left panel) at $T > T_c$ for a 
representative value of $2/9 < \mu < 1/4$. }
\end{figure} 

\section{One-Parameter Family of Generalized Symmetric Triple Well Models}

In this paper we have discussed the symmetric $\phi^{4}$-$\phi^{2}|\phi|$-$\phi^2$ triple 
well model characterized by Eq. (\ref{1}). One interesting question is if one can generalize it 
to arbitrary even power (i.e., $\phi^{4n}$-$\phi^{2n}|\phi|$-$\phi^2$) symmetric triple well 
models. The answer to the question turns out to be in the affirmative. We now discuss such  
a model and its interesting properties. In particular, we consider a symmetric model 
as given by Eq. (\ref{2a}).

\subsection{Analysis of the Generalized Model}
 
Consider the symmetric model characterized by the potential
(\ref{2a}). Let us discuss the behavior of this model for different possible
positive values of the parameters $a, b$. 

On differentiating Eq. (\ref{2a}) we obtain
\be\label{2.1}
\frac{dV}{d\phi} =  \phi \left(a - |c| |\phi|^{2n-1} + b |\phi|^{4n-2}\right)\,.
\ee
Therefore, the extrema of the potential are at 
\be\label{2.2}
\phi = 0\,,~~ |\phi|^{2n-1} = \frac{|c| \pm \sqrt{c^2-4ab}}{2b}\,.
\ee
Thus, if $c^2 < 4ab$ then the potential has minima only at $\phi = 0$. On the
other hand, if $4ab < c^2 < \frac{(2n+1)^2 ab}{2n}$, $\phi = 0$ is an 
absolute minimum while $|\phi| =|\phi_{-}| = [\frac{|c| 
+ \sqrt{c^2-4ab}}{2b}]^{1/(2n-1)}$ is a local minimum and $|\phi| = |\phi_{+}|
= [\frac{|c|-\sqrt{c^2-4ab}}{2b}]^{1/(2n-1)}$ is the maximum in between the two 
minima. The two minima at $\phi = 0$ and $|\phi| = |\phi_{-}|$ are degenerate  
with each other provided $c^2 = \frac{(2n+1)^2ab}{2n}$. Note that since it is 
$|\phi_{-}|$, one actually has three degenerate minima at $\phi = 0$ 
and $\phi = \pm |\phi_{-}|$. Finally, for $c^2 > \frac{(2n+1)^2ab}{2n}$, 
$\phi = 0$ is a local minimum and $|\phi| = |\phi_{-}|$ is an absolute minimum  
and one has a maximum at $\phi = |\phi_{+}| = [\frac{|c| - \sqrt{c^2-4ab}}{2b}]^{1/(2n+1)}$. 
Notice that this picture is very similar to that of the $\phi^2$-$\phi^{2n+2}$-$\phi^{4n+2}$ 
triple well model \cite{sck}. 

For simplicity, we now carry out a dimensional analysis so that the physics will
essentially depend on one parameter $\mu = {ab}/{c^2}$. If we write the free
energy, it is given by 
\be\label{2.3}
F_{T} = (G/2) \left(\frac{d\phi}{dx}\right)^2 + (a/2) \phi^2 - \left(|c|/(2n+1)\right) 
\phi^{2n} |\phi| +(b/4n) \phi^{4n}\,,~~a,b,c > 0\,.
\ee
On rescaling, that is considering 
\be\label{2.4}
F_{R} = \frac{c^{4n/(2n-1)}}{b^{(2n+1)/(2n-1)}}\,,~~\eta = 
\left(\frac{b}{|c|}\right)^{1/(2n-1)}\phi\,,
\ee
one finds that
\be\label{2.5}
\frac{F_{T}}{F_{R}} = g \left(\frac{d\eta}{dx}\right)^2 + (\mu/2) \eta^2 
- \left(1/(2n+1)\right)\eta^{2n} |\eta| + (1/4n) \eta^{4n}\,,
\ee
where $\mu = {ab}/{c^2}$ and $g=G/2F_R$. Thus, the rescaled potential is given by 
Eq. (\ref{2.5}) (where we relabel $\eta$ as $\phi$)
\be\label{2.6}
V(\phi) = \frac{\mu}{2}\phi^2 -\frac{1}{2n+1} \phi^{2n} |\phi| 
+\frac{1}{4n} \phi^{4n}\,.
\ee

In this case, at the point of three degenerate minima, i.e. $\mu = \frac{2n}{(2n+1)^2}$, 
one expects two (half) kink and two (half) antikink solutions. On the 
other hand, when $\mu < \frac{2n}{(2n+1)^2}$, there are two degenerate minima 
and one expects a kink (and a corresponding antikink) solution. Besides, we also 
expect pulse solutions. 

\subsection{Kink and Antikink Solutions at $\mu = \frac{2n}{(2n+1)^2}$}

We now show that at the point $\mu = \frac{2n}{(2n+1)^2}$, where there are three 
degenerate minima at $\phi = 0$ and $\phi = \pm \phi_{-} 
= [\frac{|c|+\sqrt{c^2-4ab}}{2b}]^{1/(2n-1)}$, there are two (half) kink and 
two (half) antikink solutions.

We thus need to consider the field equation
\be\label{2.7}
\phi_{xx} =  \phi^{4n-1} - \phi^{2n-1} |\phi| + \mu \phi\,.
\ee
We instead consider the following field equation
\be\label{2.8}
\phi_{xx} =  \phi^{4n-1} -  \phi^{2n} + \mu \phi
\ee
and obtain those solutions for which $\phi = |\phi|$ so that they are
automatically also the solutions of the field Eq. (\ref{2.7}). 

It is easy to check that
\be\label{2.9}
\phi = A[1 \pm \tanh(\beta x)]^{1/(2n-1)}\,,
\ee
is an exact solution of Eq. (\ref{2.8}) provided
\be\label{2.10}
A^2 = \left[\frac{n^2}{(2n+1)^2}\right]^{1/(2n-1)}\,,~~\beta^2 = 
\frac{n(2n-1)^2}{2(2n+1)^2}\,,~~\mu = \frac{2n}{(2n+1)^2}\,.
\ee
Clearly, for this solution $\phi = |\phi|$ and hence Eq. (\ref{2.9}) is also an 
exact solution of Eq. (\ref{2.7}). Observe that 
we have  both kink and antikink solutions for $\phi > 0$ as well as for
$\phi < 0$. For $\phi > 0$ we have
\be\label{2.11}
\phi = \left[\frac{n}{(2n+1)}\right]^{1/(2n-1)} [1 \pm \tanh(\beta x)]^{1/(2n-1)}\,,
\ee
while for $\phi < 0$ we have
\be\label{2.12}
\phi = -\left[\frac{n}{2n+1}\right]^{1/(2n-1)}[1 \pm \tanh(\beta x)]^{1/(2n-1)}\,,
\ee
where $\beta = \sqrt{\frac{n(2n-1)^2}{2(2n+1)^2}}$.

\subsection{Pulse Solution I}

The symmetric $\phi^{4n}$ triple well Eq.~(\ref{2.8}) also admits a hyperbolic pulse 
solution
\be\label{2.13}
\phi = \frac{A}{[B+\cosh^2(\beta x)]^{1/(2n-1)}}\,,~~B > 0\,,
\ee
provided 
\bea\label{2.14}
&&\mu = \frac{4\beta^2}{(2n-1)^2}\,,~~A = \left[\frac{(2B+1)(2n+1)\mu}
{2}\right]^{1/(2n-1)}\,, \nonumber \\
&&\frac{(2n+1)^2 \mu}{2n} = \frac{4B(B+1)}{(2B+1)^2} < 1\,.
\eea
Thus, this solution is valid only if $\mu < \frac{2n}{(2n+1)^2}$. Notice that 
for this solution $\phi = |\phi|$ and hence Eq. (\ref{2.13}) is also an exact 
solution of Eq. (\ref{2.7}). 
 
On using the well known hyperbolic identity (\ref{1.2a}) \cite{as}
we can reexpress the solution (\ref{2.13}) as a 
superposition of a kink-antikink-like solution, i.e.
\be\label{2.15}
\phi = \frac{n\mu}{{2}^{1/(2n-1)}}[\tanh(\beta x +\Delta)
-\tanh(\beta x -\Delta)]^{1/(2n-1)}\,,
\ee
where $B = \sinh^2(\Delta)$ while $A$ is as given by Eq. (\ref{2.14}).

\subsection{Pulse Solution II}

The symmetric $\phi^{4n}$ triple well Eq.~(\ref{2.8}) also admits another hyperbolic 
pulse solution
\be\label{2.16}
\phi = \left[\frac{D\cosh^2(\beta x)+DB-A}{B+\cosh^2(\beta x)}\right]^{1/(2n-1)}\,,
~~B,D > 0\,,~~n = 1, 2, 3,... \,.
\ee
In this case one finds that 
\bea\label{2.17}
&&\phi_{xx} = \frac{2\beta^2}{(2n-1)^2} \Big[-2D(2n-1)\cosh^6(\beta x)
+[3(2n-1)D+2A]\cosh^4(\beta x) \nonumber \\
&&+[2(2n-1)DB(1+B)-2(2n-1)AB -(2n+1)A]\cosh^2(\beta x) \nonumber \\ 
&&-(2n-1)B(BD-A)\Big]\,.
\eea
On substituting Eqs. (\ref{2.16}) and (\ref{2.17}) in Eq. (\ref{2.8}) and 
comparing coefficients of different powers of $\cosh^{8,6,4,2,0}(\beta x)$ 
on both sides of the equation yields
\be\label{2.18}
\mu = D -D^2\,,~~2D^2 -D = \frac{4\beta^2}{(2n-1)}\,,
\ee
\be\label{2.19}
A^2 = \frac{8B(B+1)n\beta^2}{(2n-1)^2}\,,
\ee
\be\label{2.20} 
4B(B+1)ny = 4B(B+1)(2n+1)A -(2n+1)(2B+1)Ay\,,
\ee
\be\label{2.21}
4B(B+1)n = 6(2B+1)(2n-1)A - 2(10n-7)Ay\,,
\ee
where $A = Dy$. On using Eqs. (\ref{2.20}) and (\ref{2.21}) one finds that $y$
satisfies the quadratic equation
\be\label{2.22}
2(10n-7)y^2 -(14n-5)(2B+1)y +4B(B+1)(2n+1) = 0\,,
\ee
so that
\be\label{2.23}
y = \frac{(14n-5)(2B+1) \pm \sqrt{(2B+1)^2 (14n-5)^2 - 32B(B+1)(2n+1)(10n-7)}}
{4(10n-7)}\,.
\ee
In order for the solution to be real, it is necessary that $B > y$ (note 
$A = Dy$) which is ensured if we choose the lower root for $y$ so that in 
our case $y$ is given by 
\be\label{2.24}
y = \frac{(14n-5)(2B+1) - \sqrt{(2B+1)^2 (14n-5)^2 - 32B(B+1)(2n+1)(10n-7)}}
{4(10n-7)}\,.
\ee
From Eq. (\ref{2.18}) it follows that 
\be\label{2.25}
D = \frac{1+\sqrt{1-4\mu}}{2}\,,
\ee 
so that $\mu < 1/4$, as otherwise $D$ is not real. We now show that 
$\mu > \frac{2n}{(2n+1)^2}$. On solving Eqs. (\ref{2.20}) and (\ref{2.21}) one 
can show that 
\be\label{2.26}
D = \frac{2B(B+1)n[2(10n-7)y -(2n+1)(2B+1)]}{(2n+1)y[16B(B+1)(n-1)-3(2n-1)]}\,.
\ee
Now in case $\mu = \frac{2n}{(2n+1)^2}$, it follows from Eq. (\ref{2.25}) that
$D = 2n$. On using $ D = 2n$ it easily follows that this is impossible. Note
that for $n = 1$ we have already shown that $\mu > 2/9$. Notice that 
for this solution $\phi = |\phi|$ and hence Eq. (\ref{2.16}) is also an exact 
solution of Eq. (\ref{2.7}). 
 
\section{Conclusion}

In this paper we have analyzed the symmetric $\phi^4$-$\phi^2 |\phi|$-$\phi^2$ triple well 
model given by Eq.~(\ref{1}) for various ranges of parameters. This model is relevant to 
condensed matter systems \cite{halperin} as well as for studying  tunable phase 
transitions \cite{fumika}. We then specialized it to the most interesting case when $a,b$ 
are positive and obtained kink solutions in the case when there are three degenerate minima  
as well as when there are two degenerate minima. We also obtained two pulse solutions, 
one of which is when there are two degenerate absolute minima while the other one when 
there are two local minima and one absolute minimum. In the Appendix we have given  
a few representative periodic solutions of the symmetric triple well model.  Although the 
analytic solutions are different from the standard $\phi^6$ model solutions \cite{sanati1}, 
the overall behavior is similar, albeit with a weaker nonlinearity (i.e., $\phi^4$).  A double  
well in a field ($\phi^4$-$\phi^2$-$\phi$) and the asymmetric double well in a field are 
alternative models of first order phase transitions depending on the physical situation 
\cite{sanati2, sanati3}. 

Additionally, in Sec.~IV we have generalized the symmetric $\phi^4$-$\phi^2 |\phi|$-$\phi^2$ 
model and discussed a one-parameter family of symmetric $\phi^{4n}$-$\phi^{2n} |\phi|$-$\phi^2$ 
triple well models for any positive integer $n$ and analyzed the behavior of this model 
for various ranges of parameters. In addition, we obtained kink solutions when there
are three degenerate minima and two pulse solutions, one when there are two 
degenerate absolute minima while the other in case there are two local minima 
and one absolute minimum. These findings are similar to those for the 
$\phi^{4n+2}$-$\phi^{4n}$-$\phi^2$ model \cite{sck}. 

This paper raises several open problems. Some of which are: 

\begin{enumerate}

\item For the symmetric $\phi^4$-$\phi^2 |\phi|$-$\phi^2$ model as given by Eq. 
(\ref{1}), while we have obtained a kink solution when there are two degenerate minima, 
it is not in a satisfactory form. Is it possible to obtain the kink solution in a better form?

\item Are there more hyperbolic and periodic solutions to the symmetric triple 
well model? For example (in the Appendix) we have been able to obtain the 
periodic solutions only in case $\mu = 2/9$. It would be desirable to obtain 
periodic solutions even when $\mu < 2/9$. 

\item For the generalized $\phi^{4n}$-$\phi^{2n} |\phi|$-$\phi^2$ triple well model 
as given by Eq. (\ref{2a}), can one obtain an analytic form for the kink solution when  
there are two degenerate minima? Moreover, can one obtain some periodic solutions  
for arbitrary $n$?

\item Finally, can one find physical relevance, possibly in condensed matter or field 
theory, of at least some of the generalized $\phi^{4n}$-$\phi^{2n} |\phi|$-$\phi^2$ 
models for one or two values of $n$?

\end{enumerate}

\section{Acknowledgment}

We are grateful to Fumika Suzuki for help with the figures and fruitful discussions. One of us 
(AK) is grateful to the Indian National Science Academy (INSA) for the award of INSA Honorary 
Scientist Position at Savitribai Phule Pune University. The work at Los Alamos National 
Laboratory was carried out under the auspices of the U.S. DOE and NNSA, under 
Contract No.~89233218CNA000001. 

\vskip 0.5 truecm

\section{Appendix}

We now present five periodic solutions of the symmetric $\phi^{4}$-$\phi^{2} |\phi|$-$\phi^2$ 
model as characterized by Eq. (\ref{7}) in the case $\mu = 2/9$. \\ 

{\bf Solution AI: The Periodic Kink Solution}

The symmetric $\phi^4$ triple well Eq. (\ref{8}) admits the periodic kink solution
\be\label{A1}
\phi = A + B\sqrt{m} \sn(\beta x,m)\,,
\ee
provided
\be\label{A2}
A = 1/3\,,~~\mu = 2/9\,,~~\beta = \frac{1}{3\sqrt{1+m}}\,,~~B 
= \frac{\sqrt{2}}{3\sqrt{1+m}}\,. 
\ee
Here $\sn(x,m)$ (and $\cn(x,m)$ and $\dn(x,m)$ below) are Jacobi elliptic 
functions with modulus $m$ \cite{as}. Note that the minimum value 
of $\phi$ is $A-B\sqrt{m}$ and clearly for this solution $A > B\sqrt{m}$ so
that $\phi = |\phi|$ and hence Eq. (\ref{A1}) is also an 
exact solution of Eq. (\ref{7}). 

For $\phi > 0$, we thus have periodic kink and antikink solutions given by
\be\label{A3}
\phi = \frac{1}{3}\left[1 \pm \sqrt{\frac{2m}{1+m}}\sn(\beta x, m)\right]\,,
\ee
while we also have periodic kink and antikink solutions in case $\phi < 0$
\be\label{A4}
\phi = -\frac{1}{3}\left[1 \pm \sqrt{\frac{2m}{1+m}}\sn(\beta x, m)\right]\,.
\ee
In the limit $m = 1$, the periodic 
kink solution (\ref{A1}) goes over to the hyperbolic kink solution (\ref{9}). \\ 
We hope to address some of these questions in the near future.
\begin{figure} 
\includegraphics[width=4.0 in]{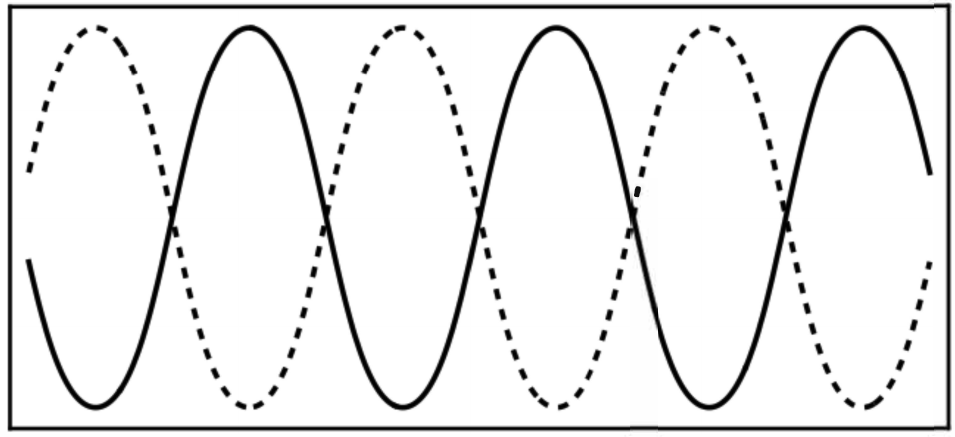}
\caption{Kink lattice solution $\phi(x)$ given by Eq. (76) corresponding to the potential 
$V(\phi)$ in the left panel of  Fig. 2 with three degenerate minima at $\mu=2/9$. 
The periodicity is $4K(m)/\beta$ antisymmetric about $x=0$.}
\end{figure}

{\bf Solution AII: Second Periodic Kink Solution}

Remarkably, the symmetric $\phi^{4}$-$\phi^{2} |\phi|$-$\phi^2$ Eq. (\ref{8}) 
admits another periodic kink solution
\be\label{A5}
\phi = F + \frac{A\sqrt{m}\sn(\beta x,m)}{D+\dn(\beta x,m)}\,,~~D > 0\,,
\ee
provided $0 < m < 1$ and further 
\be\label{A6}
F = 1/3\,,~~\mu = 2/9\,,~~D = 1\,,~~\beta = \frac{\sqrt{2}}{3\sqrt{2-m}}\,,
~~A = \frac{\sqrt{m}}{3\sqrt{2-m}}\,.
\ee
Notice that for this solution 
$F > \frac{A(1-\sqrt{1-m})}{\sqrt{m}}$ so that 
$\phi = |\phi|$ and hence Eq. (\ref{A5}) is also an 
exact solution of Eq. (\ref{7}). 

For $\phi > 0$, we thus have  periodic kink and antikink solutions given by
\be\label{A7}
\phi = \frac{1}{3}\left[1 \pm \frac{\sqrt{\frac{m^2}{2-m}}\sn(\beta x, m)}
{1+\dn(\beta x, m)}\right]\,,
\ee
while we also have periodic kink and antikink solutions in case $\phi < 0$
\be\label{A8}
\phi = -\frac{1}{3}\left[1 \pm \frac{\sqrt{\frac{m^2}{2-m}}\sn(\beta x, m)}
{1+\dn(\beta x, m)}\right]\,,
\ee
where $\beta$ is as given by Eq. (\ref{A6}). \\ 

{\bf Solution AIII: Periodic Pulse Solution}

The symmetric $\phi^{4}$-$\phi^{2} |\phi|$-$\phi^2$ Eq. (\ref{8}) also admits 
a periodic pulse solution
\be\label{A9}
\phi = A + \frac{B\sqrt{m}\cn(\beta x,m)}{\dn(\beta x,m)}\,,
\ee
provided $0 < m < 1$ and further Eq. (\ref{A2}) is satisfied. 
Notice that the minimum value of $\phi$ is $A-B\sqrt{m}$ and clearly for this solution 
$A > B\sqrt{m}$ so that $\phi = |\phi|$ and hence Eq. (\ref{A9}) is also an 
exact solution of Eq. (\ref{7}). 

For $\phi > 0$, we thus have  periodic pulse solutions given by
\be\label{A10}
\phi = \frac{1}{3}\left[1 \pm \frac{\sqrt{\frac{2m}{1+m}}\cn(\beta x, m)}
{\dn(\beta x, m)}\right]\,,
\ee
while we also have periodic pulse solutions in case $\phi < 0$
\be\label{A11}
\phi = -\frac{1}{3}\left[1 \pm \frac{\sqrt{\frac{2m}{1+m}}\cn(\beta x, m)}
{\dn(\beta x, m)}\right]\,,
\ee
where $\beta = \sqrt{{1}/{3(1+m)}}$. \\ 

{\bf Solution AIV: Second Periodic Pulse Solution}

Remarkably, the symmetric $\phi^{4}$-$\phi^{2} |\phi|$-$\phi^2$ Eq. (\ref{8}) 
also admits another periodic pulse solution
\be\label{A13}
\phi = F + \frac{A\sqrt{m}\cn(\beta x,m)}{D+\dn(\beta x,m)}\,,~~D > 0\,,
\ee
provided $0 < m < 1$ and further
\be\label{A14}
F = 1/3\,,~~\mu = 2/9\,,~~D = \sqrt{1-m}\,,~~\beta = \frac{\sqrt{2}}
{3\sqrt{2-m}}\,,~~A = \frac{\sqrt{m}}{3\sqrt{2-m}}\,.
\ee
Notice that for this solution $F > \frac{A(1-\sqrt{1-m})}{\sqrt{m}}$ so that
$\phi = |\phi|$ and hence Eq. (\ref{A13}) is also an exact solution of Eq. (\ref{7}). 

For $\phi > 0$, we thus have  periodic pulse solutions given by
\be\label{A15}
\phi = \frac{1}{3}\left[1 \pm \frac{\sqrt{\frac{m^2}{2-m}}\cn(\beta x, m)}
{\sqrt{1-m}+\dn(\beta x, m)}\right]\,,
\ee
while we also have periodic kink and antikink solutions in case $\phi < 0$
\be\label{A16}
\phi = -\frac{1}{3}\left[1 \pm \frac{\sqrt{\frac{m^2}{2-m}}\cn(\beta x, m)}
{\sqrt{1-m}+\dn(\beta x, m)}\right]\,,
\ee
where $\beta$ is as given by Eq. (\ref{A14}).  \\ 
\begin{figure} 
\includegraphics[width=4.0 in]{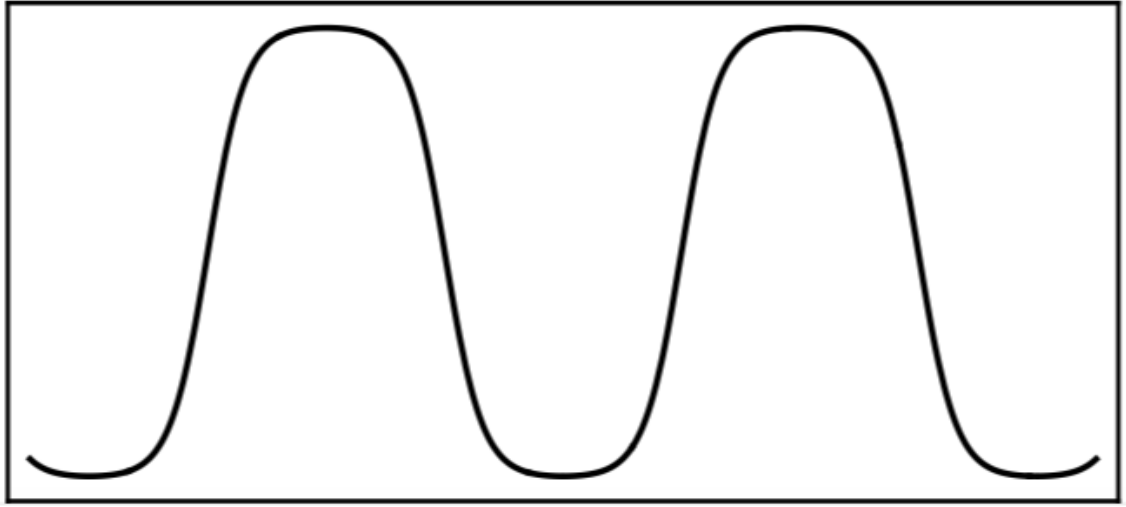}
\caption{Pulse lattice solution $\phi()$ given by Eq. (89) corresponding to the potential 
$V(\phi)$ in the left panel of  Fig. 2 with three degenerate minima at $\mu=2/9$.  
The periodicity is $4K(m)/\beta$ symmetric about $x=0$. }
\end{figure} 

{\bf Solution AV: Superposed Periodic Kink-Antikink Solution}

The symmetric $\phi^{4}$-$\phi^{2} |\phi|$-$\phi^2$ Eq. (\ref{8}) also admits 
the periodic pulse solution
\be\label{A17}
\phi = F - \frac{A\sqrt{m}\cn(\beta x,m)\dn(\beta x,m)}
{1+B\cn^2(\beta x,m)}\,,~~B > 0\,,
\ee
provided $0 < m < 1$ and further
\bea\label{A18}
&&F = 1/3\,,~~\mu = 2/9\,,~~B = \frac{\sqrt{m}}{1-\sqrt{m}}\,,~~\beta^2 = 
\frac{1}{9(1+m+6\sqrt{m}}\,, \nonumber \\
&&\beta^2 A^2 = \frac{9\sqrt{m}}{(1-\sqrt{m})^2}\,.
\eea
Observe that for this solution $F > \frac{A\sqrt{m}}{1+B}$ so that 
$\phi = |\phi|$ and hence Eq. (\ref{A17}) is also an exact solution of Eq. (\ref{7}). 

In view of the identity \cite{as} 
\be\label{A19}
\sn(y+\Delta,m)-\sn(y-\Delta,m) = \frac{2cn(y,m)\dn(y,m)\sn(\Delta,m)}
{\dn^2(\Delta,m)[1+B\cn^2(y,m)]}\,,
\ee
one can reexpress the solution (\ref{A17}) as a superposition of a periodic
kink-antikink-like solution, i.e.
\be\label{A20}
\phi = \frac{1}{3} - \frac{\sqrt{2m}}{3\sqrt{1+m+6\sqrt{m}}}
[\sn(\beta x +\Delta,m) -\sn(\beta x -\Delta,m)]\,.
\ee
For $\phi > 0$, we thus have a periodic superposed solution given by
\be\label{A21}
\phi = \frac{1}{3} - \frac{\sqrt{2m}}{3\sqrt{1+m+6\sqrt{m}}}
[\sn(\beta x +\Delta,m)-\sn(\beta x -\Delta,m)]\,,
\ee
while we also have a periodic superposed solution in case $\phi < 0$
\be\label{A22}
\phi = -\frac{1}{3} + \frac{\sqrt{2m}}{3\sqrt{1+m+6\sqrt{m}}}
[\sn(\beta x +\Delta,m) -\sn(\beta x -\Delta,m)]\,.
\ee
Here $\Delta$ is defined by $\sn(\sqrt{m}\Delta, 1/m) = \pm m^{1/4}$
where use has been made of the identity \cite{as}
\be\label{A23a}
\sqrt{m} \sn(y,m) = \sn(\sqrt{m} y, 1/m)\,.
\ee

\end{document}